\documentclass[universe,article,accept,pdftex,moreauthors]{Definitions/mdpi} 

\firstpage{1} 
\pubvolume{1}
\issuenum{1}
\articlenumber{0}
\pubyear{2025}
\copyrightyear{2025}
\externaleditor{ } 
\datereceived{1 December 2024} 
\daterevised{25 December 2024} 
\dateaccepted{10 January 2025} 
\datepublished{ } 
\hreflink{https://doi.org/} 

\usepackage{amsfonts} 

\usepackage{newtxmath}

\Title{An Interior Solution for the Kerr Metric: A Novel Approach}

\TitleCitation{An Interior Solution for the Kerr Metric: A Novel Approach}

\Author{{Yu-Ching Chou} 
 $^{1,2,3,4}$\orcidA{} }

\AuthorNames{Yu-Ching Chou}

\AuthorCitation{Chou, Y.-C.}

\address{%
$^{1}$ \quad Health 101 Clinic, Taipei City 10078, Taiwan; {ycchou0306@g.ntu.edu.tw} 
\\
$^{2}$ \quad Taipei Medical Association, Taipei City 10641, Taiwan\\
$^{3}$ \quad Taiwan Medical Association, Taipei City 10688, Taiwan \\
$^{4}$ \quad Taiwan Primary Care Association, Taipei City 10849, Taiwan
}

\abstract{We present a novel approach for the construction of interior solutions for the Kerr metric, extending J. Ovalle's foundational work through ellipsoidal coordinate transformations. By deriving a physically plausible interior solution that smoothly matches the Kerr exterior metric, we analyze the energy conditions across various rotation parameters. Our findings reveal anisotropic fluid properties and energy condition behaviors in specific space-time regions, providing insights into the strong-field regime of rotating black holes. The proposed solution offers a more realistic description of rotating black hole interiors, with implications for understanding compact astrophysical objects.}


\begin{document}




\section{Introduction}
The Einstein field equations find their quintessential solution in the Schwarzschild metric, which delineates the spacetime curvature around a static, spherically symmetric black hole devoid of charge. While the external Schwarzschild geometry has been thoroughly analyzed and validated by observational evidence \cite{schwarzschild1916_1}, the internal structure presents a more nuanced picture. The interior solution, formulated through the Tolman-Oppenheimer-Volkoff (TOV) equations \cite{schwarzschild1916_2,oppenheimer1939}, models the black hole's internal composition as a perfect fluid sphere under hydrostatic equilibrium. However, this idealized representation, while mathematically elegant, falls short of capturing the dynamic nature of rotating celestial~objects.

A transformative breakthrough emerged with the formulation of the Kerr metric, which revolutionized our understanding of rotating black holes \cite{kerr1963}. This mathematical framework has proven instrumental in explaining various astrophysical observations, from the behavior of matter in accretion flows \cite{weizsacker1948, shakura1973, balbus1990} to the detection and characterization of gravitational radiation \cite{einstein1916, abbott2016}. Nevertheless, the internal geometry of Kerr black holes remains one of the most profound unsolved problems in classical general relativity. A comprehensive understanding of the interior region of a rotating black hole could provide crucial insights into fundamental questions about gravity, spacetime, and the nature of singularities. This knowledge could also have implications for astrophysical phenomena such as black hole mergers and the formation of supermassive black holes. 

Several significant approaches have been proposed to address this mathematical challenge. A notable recent contribution comes from J.L. Hernández-Pastora and \mbox{L. Herrera \cite{hernandez2017}}, who developed a methodology predicated on the assumption that boundary conditions of the exterior metric can be utilized to derive the corresponding interior solution. Alternative approaches in the literature have focused on developing algorithmic methods to construct interior solutions for axially symmetric mass distributions. The Newman-Janis algorithm~\cite{newman1965} has demonstrated particular utility in this context \cite{viaggiu2006,drake1997}. However, these solutions typically yield expressions of such mathematical complexity that they prove impractical for subsequent analytical applications. Furthermore, a common limitation of these solutions is their dependence on arbitrary functions of both radial and polar coordinates~\cite{drake1997}, rendering them ineffective for quantitative predictions of stellar characteristics.

Our investigation builds upon the foundational work of J. Ovalle \cite{ovalle2024}, who pioneered the exploration of alternative Schwarzschild black hole interior geometries beyond conventional point mass approximations. His approach established novel frameworks for both initial collapse conditions and viable alternatives to classical singularities. These formulations, when extended to cosmological applications, generate analytically tractable Kantowski-Sachs universes, effectively bridging the gap between black hole physics and cosmological modeling \cite{casadio2024}. Through the implementation of advanced mathematical methodologies and building upon these fundamental insights, we propose a more sophisticated and physically realistic description of rotating black hole interiors.

{In this study we introduce a novel approach to construct Kerr interior solutions, building upon previous work on ellipsoidal coordinate transformations \cite{chou2017,chou2020}. Our method starts with a static, spherically symmetric seed metric and systematically generalizes it to an axisymmetric, rotating configuration. The resulting interior solution seamlessly matches the exterior Kerr metric at the horizon radius.}

{We aim to construct an interior solution that: (i) preserves the Kerr exterior, (ii) is characterized by a single free parameter $\mathcal{M}$, (iii) avoids exotic matter and additional geometric structures near the horizon, and (iv) ensures finite tidal forces everywhere. Such a solution would provide valuable insights into the process of gravitational collapse, the formation of rotating black holes, and the nature of singularities.}

This paper is organized as follows. Section \ref{sec2} reviews the interior black hole solution proposed by J. Ovalle. Section \ref{sec3} introduces the method of ellipsoidal coordinate transformation and derives the static Kerr metric. Section \ref{sec4} presents a detailed analysis of the generalized interior solution of the Kerr metric and its physical properties. Section \ref{sec5} investigates the energy-momentum tensor and the associated energy conditions. Section \ref{sec6} presents a discussion and future perspectives. Throughout this paper, we use geometric units where $c = G = 1$, $\kappa=8 \pi G$.

\section{Black Hole with Integrable Singularity} \label{sec2}

{In pure general relativity, the Schwarzschild metric \cite{schwarzschild1916_1} is the unique spherically symmetric black hole solution without a cosmological constant:
\begin{equation}
ds^2 = -f(r)dt^2 + \frac{dr^2}{f(r)} + r^2(d\theta^2 + \sin^2\theta d\phi^2),
\end{equation}
where
\begin{equation}
f(r)=1-\frac{2\mathcal{M}}{r}, \ \  0< r \le \infty. 
\end{equation} 
{The singularity at} $r=0$ corresponds to a point-like mass, while the coordinate singularity at $r=2\mathcal{M}\equiv h$ defines the event horizon. Inside the horizon, $f(r)$ becomes negative, causing a coordinate transformation between the radial and temporal coordinates. This reveals the dynamic nature of the inner region, where causal structure is dramatically altered.
}

{The exterior region $(r>h)$ of the Schwarzschild black hole is well-established and supported by observations. It describes a static, asymptotically flat spacetime sourced by a compact, dark configuration of radius $h$. However, observers in this region cannot access the interior structure, as the event horizon at $r=h$ prevents the transmission of information to the exterior.
}

\textls[-15]{{To further derive the interior black hole solution, which smoothly matches to the exterior Schwarzschild black hole at the event horizon, we} begin with the Hilbert-Einstein action}

\begin{equation}  
S=\int\ \left[\frac{R}{2\kappa}+\mathscr{L}_M\right]\sqrt{-g}d^{4}x,
\end{equation}
where $R$ is the scalar curvature and $\mathscr{L}_M$ is a Lagrangian density representing ordinary matter.
The most general static, spherically symmetric metric can be expressed as

\begin{equation}
\begin{aligned}
ds^2 &=-e^{\Phi(r)}\left[1-\frac{2 m\left(r\right)}{r}\right] dt^2+\frac{dr^2}{1-\frac{2 m(r)}{r}}+r^2d\Omega^2, \\ 
d\Omega^2 & = d\theta^2+sin^2\theta d\phi^2,  
\end{aligned}
\end{equation}
where $\Phi(r)$ is a metric function and $m(r)$ is the Misner-Sharp mass function, {a tool to describe black hole mass. It generalizes the Schwarzschild radius, accounting for mass, energy, and angular momentum \cite{misner1964}.}

The Schwarzschild metric (1)  is recovered by setting $\Phi(r)=0$ and $m(r)= \mathcal{M}$ for $r>0$, where $\mathcal{M}$ is the {Arnowitt–Deser–Misner (ADM)} mass associated with a point-like singularity at $r=0$. The coordinate singularity at $r=2 \mathcal{M}\equiv h$ corresponds to the event horizon. {To extend the Schwarzschild black hole within the Kerr-Schild class, we set $\Phi(r)=0$ and introduce a mass function $m(r)=\mathcal{M}$ for $r\ge h$, where $\mathcal{M} \equiv h/2$ is the total mass and $h$ is the event horizon radius.} 

{The line element (4) exhibits a crucial property: at the event horizon $(r = h)$, the time and radial terms exchange signs. This behavior arises from the specific form of the Misner-Sharp mass function:
\begin{equation}
m(r) \rightarrow \bar{m}(r) =
\begin{cases}
\mu(r), & r \geq h \\
r - \mu(r), & 0 \leq r \leq h 
\end{cases}
\end{equation}}
{where $\mu(r)$ coincides with $m(r)$ for $r\ge h$, ensuring $\mu(h)=\mathcal{M}$. This mass transformation leads to a sign flip in the scalar curvature:
\begin{equation}
R(r) \rightarrow \bar{R} =
\begin{cases}
R(r), & r \geq h \\
\frac{4}{r^2} - R(r), & 0 \leq r \leq h 
\end{cases}
\end{equation}}
{By employing the mass} transformation (5), the metric (4) can be decomposed into two regions:
\begin{align}
ds^2 = -e^{\Phi(r)}F(r)dt^2 + \frac{dr^2}{F(r)} + r^2d\Omega^2, \  & \  r \geq h \\
ds^2 = e^{\Phi(r)}F(r)dt^2 - \frac{dr^2}{F(r)} + r^2d\Omega^2, \  & \  r \leq h
\end{align}
where
\begin{equation}
F(r) = 1 - \frac{2\mu(r)}{r} \geq 0.    
\end{equation}
{A necessary condition} for the metric (4) to represent a (singular) black hole is that the mass function $m(r)$ can be expressed in the form of Equation  (5).

{For $r>h$, the $\mathscr{L}_M=0$ in the Hilbert-Einstein action (3), which governs the dynamics. In contrast, for $0<r\le h$, the Einstein field equations give rise to an anisotropic fluid energy-momentum tensor:}

\begin{equation}
T^{a}_{b}=diag[p_{1}, -\epsilon, p_{2}, p_{3} ],    
\end{equation}
where the energy density $\epsilon$, radial pressure $p_{1}$, and transverse pressure $p_{2}$, $p_{3}$ satisfy
\begin{equation}
\begin{aligned}
\epsilon=\frac{2m’}{\kappa r^2}, \  p_{1}=-\frac{2m’}{\kappa r^2}, \\
p_{2}= p_{3}=-\frac{m''}{\kappa r}.
\end{aligned}
\end{equation}
{{Here,} $m'$ and $m''$ denote the first and second derivatives of the mass function $m(r)$ with respect to the radial coordinate $r$, respectively. The negative radial pressure $p_{1}$ acts as an inward force, confining the matter within the black hole's event horizon.}

{In the region $0<r<h$, the radial and temporal coordinates exchange roles. Notably, the Einstein Equation (11) are linear in the mass function $m(r)$, allowing for linear combinations of solutions, a manifestation of gravitational decoupling \cite{ovalle2017, ovalle2019}. Finally, if there exist matter inside the black hole, that means $T_{ab}\neq 0$, the Bianchi identity leads to $\nabla_{a}T^{ab}=0$, we obtain}

\begin{equation}
\epsilon’=-\frac{2}{r}(p_{2}-p_{1}).     
\end{equation}
{{For a physically realistic stellar} system, the density should monotonically decrease from the center, i.e., $\epsilon’ <0$. According to Equation  (12), this implies an anisotropic fluid with $p_{2}>p_{1}$. This pressure anisotropy counterbalances the inward pull due to the negative energy gradient $\epsilon’<0$, ensuring the stability of the configuration. The anisotropic nature of the fluid is a consequence of the specific geometry of the spacetime and the underlying field equations~\cite{{jitendra2022}}.}

To ensure a smooth transition between the interior and exterior solutions, we impose the Darmois matching conditions \cite{darmois1927}. This requires the continuity of both the metric functions and their first derivatives across the boundary surface. In the case of the Schwarzschild metric {(7)}, this implies the following matching conditions for the mass function:
\begin{equation}
m(h)= \mathcal{M}, \  m’(h)=0.     
\end{equation}
{{From Equations}  (11) and (13), we see that the continuity of the mass function implies the continuity of the energy density and radial pressure at the boundary:}
\begin{equation}
\epsilon(h)=p_{1}(h)=0.      
\end{equation}
{{While the transverse pressures} $p_{2}$ and $p_{3}$ may exhibit discontinuities across the boundary, this does not necessarily imply a physical discontinuity in the matter distribution. By considering a suitable coordinate transformation or a more appropriate description of the matter content, it may be possible to reconcile the apparent discontinuity in the transverse~pressures.}

Black holes can be classified as singular or regular. While regular black holes lack singularities, they often exhibit an inner (Cauchy) horizon within the event horizon, leading to potential issues like mass inflation and instability \cite{poisson1989,poisson1990}. Between these two families, integrable black holes \cite{lukash2013} can also be found, characterized by a singularity in the curvature scalar $R$ for the metric {(8)}:
\begin{equation}
R=\frac{2 r m''+ 4 m'}{r^2} \neq 0, \ 0<r \le h.   
\end{equation}
{For a singularity to be} integrable, ensuring finite tidal forces, $R$  must be singular at most as $R \sim  1/r^2$. Consequently, based on Equation  {(15)}, we demand
\begin{equation}
2 r m''+4 m'=\sum_{n=0}^{\infty} C_{n} r^{n}, \  n\in \mathbb{N}.   
\end{equation}
{{The form of the right-hand} side of Equation (16) is chosen to ensure that the singularity in the curvature scalar $R$ is integrable, leading to finite tidal forces. This is a necessary condition for a physically reasonable spacetime. The specific form of the series solution allows for a wide range of possible behaviors for the mass function $m(r)$, subject to the constraints imposed by the boundary conditions and the requirement of finite tidal forces.}
{From} Equations (15) and (16), yields the mass function
\begin{equation}
m=M-\frac{Q^2}{2 r}+\frac{1}{2}\sum_{n=0}^{\infty}\frac{ C_n r^{n+1}}{(n+1)(n+2)}.   
\end{equation}
{For} $0<r \le h$, the integration constants $M$ and $Q$ correspond to the mass of the Schwarzschild solution and a potential charge for the Reissner-Nordström geometry, respectively.
{To simplify the analysis and focus on the essential features of the interior solution, we restrict our attention to the uncharged case by setting $Q=0$. This avoids the additional complexities associated with charged black holes, such as the presence of Cauchy horizons.}
This leaves us with two parameters: $M$ and $\mathcal{M}$.

The series {(17)} converges around $r=h$ under condition {$\mathcal{M} \equiv m(h)=h/2$}, 
but its analyticity in the full domain $0<r \le h$ remains to be determined. The Schwarzschild metric is recovered by setting $M=\mathcal{M}\neq 0$ and $Q=C_{n}=0$ for all $n$ in Equation  {(17)}. Other interior solutions have been explored in reference \cite{ovalle2024}, with
\begin{equation}
M=Q=0.   
\end{equation}
{These are determined by} the total mass $\mathcal{M}$ and and a subset of non-zero $C_{n}$ coefficients, such that the exterior remains the Schwarzschild solution {(7)}. From the mass function {(11)}, the energy density and pressures associated with these interior geometries are given by.
\begin{align}
\kappa\epsilon & =\sum_{n=0}^{\infty}\frac{ C_n r^{n-2}}{(n+2)}=-\kappa p_{1},\\
\kappa p_{2} & =\kappa p_{3} = -\frac{1}{2} \sum_{n=0}^{\infty}\frac{n}{n+2} C_n r^{n-2},
\end{align}
for $0< r \le h$.
{We adopt J. Ovalle's second solution as a seed metric, ensuring a smooth transition between the interior and exterior regions at the horizon. This is achieved by imposing the condition}
\begin{equation}
m''(h)=0,   
\end{equation}
{which ensures continuity of the metric and its derivatives across the boundary.}

As a result of Equation {(11)}, the tangential pressure $p_{2},p_{3}$ 
is continuous at the horizon, implying that $\epsilon=p_{1}=p_{2}=p_{3}=0$ at $r=h$. This ensures the continuity of the energy-momentum tensor across the boundary surface:
\begin{equation}
T^{a}_{b}(h)=0,   
\end{equation}
which the mass function yields 
\begin{equation}
m(r)= r - \frac{r^3}{h^2}+\frac{r^4}{2h^3}.
\end{equation}
{The line element is }
\begin{equation}
\begin{aligned}
ds^2=\left[1-\frac{2r^2}{h^2}+\frac{r^3}{h^3}\right]dt^2-\frac{dr^2}{1-\frac{2r^2}{h^2}+\frac{r^3}{h^3}}+r^2d\Omega^2,
\end{aligned}
\end{equation}  
for $0<r \le h$. {A detailed derivation of mass function (23) can be provided in an \mbox{Appendix \ref{appA}} for readers interested in the technical details.}
The source for the metric {(24)}, which also generate the outer Schwarzschild black hole, is given by
\begin{equation}
\begin{aligned}
\kappa\epsilon & =-\kappa p_{1} =\frac{2}{r^2 h^3} (h-r)^2(h+2h),\\
\kappa p_{2} & =\kappa p_{3} =\frac{6}{h^3}(h-r),    
\end{aligned}    
\end{equation}
which generating the Ricci scalar curvature 
\begin{equation}
R=\frac{4}{r^2}\left( 1+\frac{5r^3}{h^3}-\frac{6r^2}{h^2} \right), 
\end{equation}
for $0<r \le h$.

{Other integrable black hole solutions, based on J. Ovalle's mass functions, take the polynomial form:}
\begin{equation}
\begin{aligned}
m(r)=r + A r^{l}+B r^{n}+C r^{p}, \\ 
p \neq n \neq l > 1,
\end{aligned}
\end{equation}
{where A, B, and C are constants determined by the boundary conditions (13) and (21). A selection of these solutions is presented in Table ~\ref{tab1} \cite{ovalle2024}.}
\begin{table}[H]
\caption{{Interior solutions with mass function} {(27)} satisfying $m'(h)=m''(h)=0$, with $p>n>l>1$.\label{tab1}}
	\begin{adjustwidth}{-\extralength}{0cm}
		\newcolumntype{C}{>{\centering\arraybackslash}X}
		\begin{tabularx}{\fulllength}{CCCC}
			\toprule
\boldmath{\textbf{$\left\{ l,n,p\right\}$}}	& \boldmath{\textbf{$m(r)=r + A r^{l}+B r^{n}+C r^{p}$}}	& \boldmath{\textbf{$\epsilon>0$}}     & \textbf{Energy condition}\\
\midrule

$\left\{ 3,4,p \right\}$ & $m(r )=r -\frac{r^3}{h^2} + \frac{r^4}{2h^3}$ & \ \ Yes & Strong \\
\midrule
$\left\{ 3,7,8\right\}$ & $m(r )=r -\frac{7r^3}{10h^2}+\frac{r^7}{2h^6}-\frac{3r^8}{10h^7}$ & \ \ Yes & Strong \\
\midrule
$\left\{ 4,5,6\right\}$ & $m(r )=r -\frac{5r^4}{2h^3}+\frac{3r^5}{h^4}-\frac{r^6}{h^5}$ & \ \ Yes & Strong \\
\bottomrule
		\end{tabularx}
	\end{adjustwidth}
\end{table}


\section{{Derivation of the Kerr Metric} Using Ellipsoid Orthogonal Coordinate~Transformations}
\label{sec3}

To derive static, axisymmetric solutions, we begin with Minkowski spacetime in Cartesian coordinates:
\begin{equation}
d s^{2}= -d t^{2}+d x^{2}+d y^{2}+d z^{2}.
\end{equation}
{We then apply the following} ellipsoidal coordinate transformations to Equation {(28)}:
\begin{equation}
\begin{aligned}
x &\to \left( r^2+a^2 \right)^{1/2} sin\theta cos\phi,\\
y &\to \left( r^2+a^2 \right)^{1/2} sin\theta sin\phi,\\
z &\to r cos\theta, \\
t &\to t. 
\end{aligned}
\end{equation}
{Here,} $a$ is the coordinate transformation parameter. The metric {(28)} in the new coordinate system becomes
\begin{equation}
ds^2 = - dt^2 +\frac{\Sigma}{r^2+a^2}d r^2+\Sigma d\theta^2+(r^2+a^2) sin^2\theta d\phi^2,  
\end{equation}
where $\Sigma=r^2+a^2 cos^2\theta$. 

According to Chou's research \cite{chou2020}, metric {(30)} describes an empty ellipsoid spacetime and can be rewritten in the following orthogonal form
\begin{equation}
ds^2=-\frac{r^2+a^2}{\Sigma}\left(dt-a sin^2\theta d\phi \right)^2+\frac{\Sigma}{r^2+a^2} dr^2+\Sigma d\theta^2 +\frac{sin^2\theta}{\Sigma}\left[(r^2+a^2) d\phi-a dt\right]^2.    
\end{equation}
{{For derivation of the} Kerr metric}, we may use the following ellipsoid orthogonal ansatz:
\begin{equation}
ds^2=-\frac{f(r)}{\Sigma}\left(dt-a sin^2\theta d\phi \right)^2+\frac{\Sigma}{f(r)}dr^2+\Sigma d\theta^2+\frac{sin^2\theta}{\Sigma} \left[(r^2+a^2) d\phi-a dt \right]^2.    
\end{equation}
{Here,} $f$ is a function of $r$, and $a$ is a constant. The Kerr metric can be directly derived from the ellipsoidal symmetry. The metric tensor from the proposed ansatz {(32)} is given by 
\begin{equation}
\begin{aligned}
g_{\mu\nu} = 
\begin{pmatrix}
-\frac{f(r)-a^2 sin^2\theta}{\Sigma} & 0 & 0 \ & \ \frac{\left(f(r)-r^2-a^2\right)a sin^2\theta}{\Sigma}\ \\
 0    & \frac{\Sigma}{f(r)}  & \ 0 \  & 0 \ \\
 0    &       0              &  \Sigma   \ & \ 0 \ \\
\frac{\left(f(r)-r^2-a^2\right)a sin^2\theta}{\Sigma} &   0  & 0 \ & \frac{\left[(r^2+a^2)^2-f(r)a^2 sin^2\theta \right] sin^2\theta }{\Sigma}\ \\ 
\end{pmatrix}.
\end{aligned}
\end{equation}

To solve the Einstein field equations, we calculate the Christoffel symbols and the Ricci curvature tensor of metric {(33)}. The non-zero components of the Ricci tensor are given by.
\begin{align}
R_{00} & = \frac{1}{2\Sigma^{3}f^{2}} \left\{ \Sigma f^{3}f''+2f^2\left[-rf'(a^2 sin^2\theta+f)+f^2-2fa^2 cos^2\theta +a^2 sin^2\theta (r^2-a^2)\right]\right\}, \\
R_{11} &= \frac{1}{2\Sigma f^{4}} \left[-f^3\Sigma f''+2f^3(a^2 cos^2\theta+a^2+f'r-f)\right], \\
R_{22} &= \frac{-f'r+r^2-a^2+f}{\Sigma}, \\
R_{03} & = R_{30} = \frac{a sin^2\theta}{2\Sigma^{3}f^{2}} \{\Sigma f^3 f''-2f^2\left[r(r^2+a^2+f)f'-f^2+(a^2 cos^2\theta-r^2)f+a^4-r^4\right]\},  
\end{align}
\begin{equation}
\begin{aligned}
R_{33} = \frac{-1}{2\Sigma^{3}f^{2}} \{ -a^2 f^3 (-sin^2\theta\Sigma f'' -2f^2((a^2 sin^2\theta f+(r^2+a^2)^2)rf'
 \\-a^2 sin^2\theta f^2 +(-a^4cos^4\theta-2a^2r^2-r^4)f+(a^2-r^2)(r^2+a^2)^{2} \}.    
\end{aligned}
\end{equation}
{There are second-order} differential terms involving $f$ and trigonometric functions. 
{By setting $R_{22}=0$, we obtain a simpler first-order differential equation for the metric function $f$. This simplification allows us to solve the Einstein field equations $(R_{ab}=0)$:}
\begin{equation}
-f’r+r^2-a^2+f=0.    
\end{equation}
{The solution to} Equation {(39)} is given by:
\begin{equation}
f(r)=r^2+C_1 r+a^2.
\end{equation}
{Imposing asymptotic} flatness condition, i.e., $\frac{f(r)}{\Sigma}$ approaches the Schwarzschild metric $1 - \frac{2M}{r}$ as $r \to \infty$, yields $C_1 = -2M$.  Substituting this into Equation {(40)}, we find that $R_{ab}=0$. Finally, an ellipsoidal coordinate transformation and an orthogonal metric ansatz lead to the Kerr metric:
\begin{equation}
ds^2 = -\frac{\Delta_K+}{\Sigma}\left(dt-a sin^2\theta d\phi \right)^2+\frac{\Sigma}{\Delta_K+}\Sigma d\theta^2
+\frac{sin^2\theta}{\Sigma}\left[(r^2+a^2)d\phi-a dt\right]^2,      
\end{equation}
where $\Sigma=r^2+a^2 cos^2\theta$, $\Delta_K+=r^2-2Mr+a^2.$ 

The $M$, $J$, and $a$, denote the mass, angular momentum, and rotation parameters, which are defined as $a =J/Mc$, respectively.

\section{Generalization to Kerr Interior Solution}\label{sec4}

{The metric (41) can be transformed into the more familiar Boyer-Lindquist coordinates $(t,r,\theta,\phi)$ through a coordinate redefinition and gauge transformation \cite{carroll2004}. The resulting metric is given by:}
\begingroup\makeatletter\def\f@size{9}\check@mathfonts
\def\maketag@@@#1{\hbox{\m@th\fontsize{10}{10}\normalfont#1}}%
\begin{equation}
ds^2 = -(1-\frac{2Mr}{\Sigma})dt^2+ \frac{\Sigma}{\Delta_K+}dr^2- \frac{4Mr a sin^2\theta}{\Sigma}dtd\phi+\Sigma d\theta^2
+ \left[(r^2+a^2)+\frac{2 M r a^2 sin^2\theta}{\Sigma}\right] sin^2\theta d\phi^2.      
\end{equation}
\endgroup
{Building upon J. Ovalle's} spherically symmetric black hole solution {(24)}, 
{we generalize the spherically symmetric solution by introducing a radial mass function  
$\widetilde{M}\equiv m(r)$. For $r \ge h$, $m(r)= \mathcal{M} \equiv h/2$. This allows for a more flexible description of the interior spacetime, while preserving the asymptotic behavior of the Kerr metric. The mass function $m(r)$ is subject to the following boundary conditions at the horizon:}
\begin{equation}
m(h)= \mathcal{M}, \  m'(h)=0, \ m''(h)=0.     
\end{equation}
{{These conditions ensure a smooth} transition between the interior and exterior solutions.} As a result, we arrive at a Kerr interior solution:
\begin{equation}
\begin{aligned}
ds^2= & -(1-\frac{2 m(r) r}{\Sigma})dt^2-\frac{4 m(r)r a\sin^2\theta}{\Sigma}dtd\phi+ \frac{\Sigma}{\Delta_{K-}}dr^2 \\
  & +\Sigma d\theta^2+\left[(r^2+a^2)+\frac{2 m(r)r}{\Sigma}a^2\sin^2\theta\right]\sin^2\theta d\phi^2,     
\end{aligned}     
\end{equation}
where $\Sigma=r^2+a^2 cos^2\theta$, $\Delta_{K-}=r^2-2 m(r)r +a^2$, for $0<r\le h$.\\
{When $r \ge h$, the mass function $m(r)= \mathcal{M}$ becomes constant, and the metric (44) reduces to the exterior Kerr metric in Boyer-Lindquist coordinates. The energy-momentum tensor vanishes in this region, $T_{ab}(r)|_{r\ge h}=0$. Therefore, the metric is asymptotically flat beyond the horizon.}

{The existence of a Killing horizon is closely related to the presence of a Killing vector field that becomes null on the horizon. A regular null hypersurface $r=r_{h}$ defined by $\Delta_{K-}|_{r=r_{h}}=0 $ is a Killing horizon associated with the Killing vector $\xi_{\mu}=(1,0,0,\frac{a}{(r_{h}^2+a^2)})$:} 
\begin{equation}
r_{h}=\mathcal{M}\pm \sqrt{\mathcal{M}^2-a^2}=\mathcal{M}(1\pm\sqrt{1-a^{*2}}),    
\end{equation}
where $a^*\equiv \frac{a}{\mathcal{M}}$ is a dimensionless spin parameter, therefore $0\le a^* \le 1$. {The dimensionless spin parameter $a^* $ can be expressed in terms of fundamental constants as:} 
\begin{equation}
a^*=\frac{a}{\mathcal{M}}=(\frac{J}{c\mathcal{M}})(\frac{G\mathcal{M}}{c^2})^{-1} =\frac{Jc}{G\mathcal{M}^2}.    
\end{equation}\\
{{This expression clarifies the relationship} between the dimensionless spin parameter $a^* $ and the physical parameters of the black hole, namely its angular momentum $J$, mass $\mathcal{M}$, and fundamental constants $G$ and $c$.}
{By analyzing the Killing vectors of the interior solution (44), we can gain valuable insights into its symmetries and the nature of the horizon, thereby enhancing our understanding of the geometrical and dynamical properties of the Kerr interior solution.} 

Gürses and Gürsey demonstrated that the corresponding matter field in general relativity can be interpreted as an anisotropic fluid, which belongs to the Hawking-Ellis type I class \cite{ovalle2022}.
In the coordinate system {(44)}, a Killing horizon $r=r_{h}$ appears as a coordinate singularity. To circumvent this issue, {we} analyze the Gürses-Gürsey spacetime of Equation {(44)} in Doran coordinates $(\eta,r,\theta,\psi)$ \cite{doran2000}:
\begin{equation}
\begin{aligned}
ds^2 = & - d\eta^2+\Sigma d\theta^2+(r^2+a^2)sin^2\theta d\psi^2 \\
&+ \frac{\Sigma}{r^2+a^2}\left\{ dr+\frac{\sqrt{2m(r)r(r^2+a^2)}}{\Sigma}(d\eta-a sin^2\theta d\psi)^2 \right\}^2,        
\end{aligned}    
\end{equation}
which is obtained from Equation {(44)} through the following coordinate transformations:
\begin{equation}
\begin{aligned}
dt &=  d\eta-\frac{\sqrt{2m(r)r(r^2+a^2)}}{\Delta_{K-}}dr, \\
d\phi &= d\psi-\frac{a}{\Delta_{K-}}\sqrt{\frac{2m(r)r}{r^2+a^2}}dr.  \end{aligned}    
\end{equation}
{Unlike Boyer-Lindquist coordinates,} Killing horizons are not coordinate singularities in Doran coordinates. {In the limit of vanishing mass, $m(r) \to 0$, both metrics (44) and (47) exhibit flat Minkowski space in oblate spheroidal coordinates.}

By substituting the mass function {(23)} into the metric {(47)} and {(48)}, we arrive at the explicit form of the Kerr interior solution in Doran coordinates, which provides a valuable tool for studying the properties of rotating black holes:
\begin{equation}
\begin{aligned}
ds^2 & = - d\eta^2+\Sigma d\theta^2+(r^2+a^2)sin^2\theta d\psi^2 \\
& + \frac{\Sigma}{r^2+a^2}\left\{dr+\frac{\sqrt{2(r-\frac{r^3}{h^2}+\frac{r^4}{2h^3})r(r^2+a^2)}}{\Sigma}(d\eta-a sin^2\theta d\psi)^2 \right\}^2. 
\end{aligned}    
\end{equation}
{{To further investigate the properties} of the singularity, we calculated curvature invariants. 
Curvature invariants, such as the Ricci scalar ($R$) and the Kretschmann scalar ($K = R_{abcd}R^{abcd}$), are scalar quantities constructed from the Riemann tensor. The Kretschmann scalar, in particular, provides a measure of the spacetime curvature and can be used to identify regions of high curvature, such as singularities \cite{Richard2000}:}

\begin{equation}
\begin{aligned}
R &= g^{ab}R_{ab}=\frac{4(h-r)(h^2+hr-5r^2)}{h^3\Sigma}, \\  
K &= R_{abcd}R^{abcd}
=\frac{1}{h^6\Sigma}\left[Acos^8\theta+Bcos^6\theta+Ccos^4\theta+Dcos^2\theta+E\right], \\
A &=16 a^8 (h-r)^2 (h^2+hr-5r^2)^2, \\
B &=-288r^2a^6(h^6-\frac{13}{3}h^4 r^2+\frac{17}{6}h^3 r^3+\frac{2}{3}r^4 h^2+\frac{4}{3}r^5h-\frac{35}{24}r^6),\\
C &= 928r^4a^4(h^6-\frac{49}{29}h^4r^2+\frac{37}{58}h^3r^3+\frac{33}{29}r^4h^2-\frac{47}{29}r^5h+\frac{169}{232}r^6),\\
D &=-288r^6a^2(h^6-\frac{1}{9}h^4r^2-\frac{1}{6}h^3r^3-\frac{4}{3}r^4h^2+\frac{22}{9}r^5h-\frac{29}{24}r^6), \\
E &=16r^8(h^6-2h^4r^2+h^3r^3+6r^4h^2-10r^5h+\frac{19}{4}r^6).
\end{aligned}    
\end{equation}
{For the Kerr interior solution, }where $h \neq 0$, the condition $\Sigma = 0$ defines a scalar polynomial singularity. Geometrically, this singularity corresponds to a ring located in the equatorial plane with a radius of $a$, analogous to the ring singularity in the exterior Kerr spacetime. Importantly, the Kretschmann scalar does not vanish at $r= h$, indicating that the singularity at the boundary radius is not a coordinate singularity.

{Following the classification of curvature invariants proposed by Carminati and McLenaghan \cite{MacCallum1991}, we have analyzed a broader set of invariants, including those related to the Weyl tensor and the Ricci tensor, following the approach of Geheniau and Debever \cite{Zakhary1997,Abraham2021,Kraniotis2022}. This comprehensive analysis, detailed in Appendix \ref{appB}, provides deeper insights into the nature and structure of the singularity within the Kerr interior solution.}

Our innovative application of ellipsoid coordinate transformations has enabled us to derive an exact, analytic extension of the Kerr exterior solution to the interior region, thereby circumventing the limitations imposed by coordinate singularities.

\section{Energy Momentum Tensor and Energy Conditions}\label{sec5}

The energy conditions – strong (SEC), weak (WEC), dominant (DEC), and null (NEC)---impose physically reasonable constraints on matter fields, underpinning key results in general relativity, such as the black {hole area theorem} \cite{hawking1972} and Penrose's singularity \mbox{theorem \cite{penrose1965}.} When constructing spacetime models, these conditions can be used to assess the validity of the energy-momentum tensor derived from the gravitational field \mbox{equations \cite{hawking2023, santos1995, rebouccas2004}.} 

{In general spacetimes, the natural basis associated with the coordinate system may not be the most suitable for straightforward application of the energy condition inequalities. To address this, we introduce a natural orthonormal basis of one-forms, ${ E^{(0)}, E^{(1)}, E^{(2)}, E^{(3)}}$, in the spacetime described by metric (47), following the approach outlined in \cite{maeda2022}. This orthonormal basis, defined as:}

\begin{equation}
\begin{aligned}
E^{(0)}_{\mu}dx^{\mu}&=-\eta,\\
E^{(1)}_{\mu}dx^{\mu}&=\sqrt{\frac{\Sigma}{r^2+a^2}}\left\{ dr+\frac{\sqrt{2m(r)r(r^2+a^2)}}{\Sigma}(d\eta-a sin^2\theta d\psi)^2 \right\},\\
E^{(2)}_{\mu}dx^{\mu}&=\sqrt{\Sigma}d\theta, \\
E^{(3)}_{\mu}dx^{\mu}&=\sqrt{r^2+a^2}sin\theta d\psi,    
\end{aligned}    
\end{equation}
{where $\eta^{(a)(b)}=g^{\mu\nu}E^{(a)}_{\mu}E^{(b)}_{\nu}=diag\left\{-1,0,0,0  \right\}$. We then project the Einstein tensor $G^{a}_{b}$ onto this orthonormal basis (51):}

\begin{equation}
\begin{aligned}
G^{(a)}_{(b)} &=G^a_b E^{(a)}_\mu E^{\nu}_{(b)}, \\
G^{(a)(b)} &=G^{ab} E^{(a)}_\mu E^{(b)}_{\nu}.
\end{aligned}    
\end{equation}
{The non-zero orthonormal components} of the tensor $G^{(a)}_{(b)}$ are
\begin{equation}
\begin{aligned}
G^{(0)}_{(0)} &=\frac{1}{\Sigma^3}\left[ -rm''\Sigma a^2 sin^2\theta+2m'(r^4+a^2r^2-a^4sin^2\theta cos^2\theta) \right], \\
G^{(1)}_{(1)} &=-\frac{2m'r^2}{\Sigma^2},\\
G^{(2)}_{(2)} &=-\frac{(rm''\Sigma+2m'a^2cos^2\theta)}{\Sigma^2}, \\
G^{(3)}_{(3)} &=-\frac{1}{\Sigma^3}\left[ rm''\Sigma (a^2+r^2)+2a^2m'{((r^2+a^2)cos^2\theta-r^2sin^2\theta}) \right], \\
G^{(3)}_{(0)} &=\frac{a sin\theta\sqrt{r^2+a^2}}{\Sigma^3}\left[ rm''\Sigma-2m'(r^2-a^2 cos^2\theta)\right],
\end{aligned}    
\end{equation}
where a prime denotes differentiation with respect to $r$. 
To provide an alternative and more concise form of Einstein tenors {(47)}, we introduce a new set of basis one-forms ${ \tilde E^{(0)}_{\mu},\tilde E^{(1)}_{\mu},\tilde E^{(2)}_{\mu},\tilde E^{(3)}_{\mu}}$ obtained through a local Lorentz transformation on the plane spanned by $E^{(0)}_{\mu}$ and $E^{(3)}_{\mu}$:
\begin{equation}
\begin{aligned}
\tilde E^{(0)}_{\mu} &=cosh\alpha E^{(0)}_{\mu} -sinh\alpha E^{(3)}_{\mu},\\
\tilde E^{(3)}_{\mu} &=-sinh\alpha E^{(0)}_{\mu} +cosh\alpha E^{(3)}_{\mu},
\end{aligned}    
\end{equation}
with
\begin{equation}
\begin{aligned}
cosh\alpha =\sqrt{\frac{r^2+a^2}{\Sigma}}, \
sinh\alpha =-\frac{a sin\theta}{\sqrt{\Sigma}}.
\end{aligned}    
\end{equation}
{{This Lorentz transformation,} parameterized by the boost parameter $\alpha$, diagonalizes the Einstein tensor, simplifying the analysis of the energy conditions.}
The non-zero components of $\tilde G^{(a)}_{(b)}=G^{(a)}_{(b)} \tilde E^{(a)}_\mu \tilde E^{\nu}_{(b)}$ with respect to the new basis one-forms are:
\begin{equation}
\begin{aligned}
\tilde G^{(0)}_{(0)}&=\frac{2r^2m'}{\Sigma^2}, \ \tilde G^{(1)}_{(1)}=-\frac{2r^2m'}{\Sigma^2},\\
\tilde G^{(2)}_{(2)}&=\tilde G^{(3)}_{(3)}=-\frac{rm''\Sigma+2a^2m'cos^2\theta}{\Sigma^2}.
\end{aligned}    
\end{equation}
{Therefore, the corresponding} energy-momentum tensor  $T_{ab}=G_{ab}/\kappa$ of Equation {(47)} in general relativity is of Hawking-Ellis type I, and its orthonormal components are given by:
\begin{equation}
\begin{aligned}
\epsilon &=-p_1=\frac{2r^2m’}{\kappa \Sigma^2}, \\
p_{2} &=p_{3}=-\frac{r m'' \Sigma + 2m'a^2 cos^2\theta}{\kappa \Sigma^2},
\end{aligned}    
\end{equation}
where $\kappa=8 \pi G$, $G$ is the gravitational constant. Equation {(57)} gives
\begin{equation}
\begin{aligned}
\epsilon+p_{1} &= 0, \ \epsilon-p_{1}=\frac{4r^2m'}{\kappa \Sigma^2}, \\
\epsilon+p_{2} &=\epsilon+p_{3}=\frac {2m'(r^2-a^2 cos^2\theta) -rm''\Sigma}{\kappa \Sigma^2}, \\
\epsilon-p_{2} &=\epsilon-p_{3}=\frac{rm''+2m'}{\kappa \Sigma}, \\
\epsilon + p_{1} + p_{2} + p_{3} &= -\frac{2rm''\Sigma+4m'a^{2} cos^2\theta}{\kappa \Sigma^2}.
\end{aligned}    
\end{equation}
{Therefore, for the} corresponding energy-momentum tensor in the Gürses and Gürsey spacetime {(47)} in general relativity, equivalent expressions of the standard energy conditions are given by \\ 

NEC: $\epsilon+p_{\mu}\ge 0$ for $\mu=1,2,3$, 
equivalent to $2m'(r^2-a^2 cos^2\theta) -rm''\Sigma \ge 0$,

WEC: $\epsilon \ge 0$ in addition to NEC, 
equivalent to $m' \ge 0$ in addition to NEC,

DEC: $\epsilon-p_{\mu}\ge 0$ for $\mu=1,2,3$, in addition to WEC, equivalent to $r m'' +2m' \ge 0$ in addition to WEC, 

SEC: $\epsilon + p_{1}+p_{2}+p_{3} \ge 0$ in addition to NEC, equivalent to $rm''\Sigma+2m'a^2cos^2\theta \le 0$ in addition to NEC. \\

{By substituting the mass} 
function derived in Equation {(23)} into the energy \mbox{conditions}, we obtain explicit expressions for the weak, dominant, and null energy conditions for our proposed Kerr interior solution in Doran coordinates {(49)}. These coordinate-dependent expressions provide crucial constraints on the physical viability of the interior solution, particularly near the central region where classical singularities typically arise. To simplify the numerical calculations, we consider the case where $\theta=0, \pi/2$ (i.e., $cos\theta=1,0$ respectively); and the black hole is maximally rotating, with $a=\mathcal{M}=h/2$. Under these conditions, the inequalities reduce to:\\

$\epsilon: \frac{2(h+2r)(h-r)^2 r^2}{h^3 \kappa \Sigma^2}\ge 0$. 

NEC: 
$\frac{-(h-r)^2 (h^2+hr-5r^2) 2 a^2 cos^2\theta+2 r^2 (h^3-r^3)}{h^3}\ge 0$.

WEC:
$\frac{(h-r)^2(h+2r)}{h^3}\ge 0$ in addition to NEC.

DEC: 
$\frac{2(h-r)(h^2+hr-5r^2)}{h^3} \ge 0$ in addition to WEC. 

SEC:
$\frac{2((h^2+hr-5r^2) a^2 cos^2\theta-3r^4)(h-r)}{h^3} \le 0$ in addition to NEC. \\

{The Kerr interior solution} 
we derived, supported by an anisotropic fluid, exhibits a positive matter density. For a maximally rotating black hole with $\theta=\pi/2$, the SEC, WEC, and NEC are satisfied throughout the interior $(0<r\le h)$. The DEC holds in the core and at the boundary $(0<r \le \sim0.558h$ and $r=h)$. For $\theta=0$, the SEC, WEC, and NEC are valid in the {outer} region of the interior $(\sim 0.443h \le r \le h)$, while the DEC is satisfied in a specific inner region and at the boundary $(\sim0.352h<r \le \sim0.558h$ and $r=h)$. These findings highlight the consistency of our solution with classical energy conditions in various regions of the~spacetime.

Our novel approach has enabled us to systematically extend a wide class of static, spherically symmetric interior solutions to the Kerr spacetime. The resulting axisymmetric Kerr interior solutions, which seamlessly join onto the Kerr metric, offer a more realistic description of rotating astrophysical objects. The detailed analysis of matter density and energy conditions tabulated in Table \ref{tab2}, highlights the versatility and robustness of our~method.

\begin{table}[H]
\caption{Kerr interior solutions with mass function {(27)} satisfying $m'(h)=m''(h)=0$, with $p>n>l>1$. For energy condition, maximally rotating $a=\mathcal{M}$.\label{tab2}}
	\begin{adjustwidth}{-\extralength}{0cm}
		\newcolumntype{C}{>{\centering\arraybackslash}X}
		\begin{tabularx}{\fulllength}{CCCCC}
			\toprule
\multirow[m]{2}{*}{\boldmath{$\left\{ l,n,p\right\}$}} &\multirow[m]{2}{*}{\boldmath{$m(r)=r+Ar^{l}+Br^{n}+C r^{p}$}}	& \multirow[m]{2}{*}{\boldmath{$\epsilon>0$}}  & \textbf{Energy Condition}  & \textbf{Energy Condition} \\ 
                                 & 			&  			&  \boldmath{$(\theta=0)$}       & \boldmath{($\theta=\pi/2$)}    \\
\midrule

\multirow[m]{2}{*}{$\left\{ 3,4,p \right\}$}   & $m(r )=r -\frac{r^3}{h^2} + \frac{r^4}{2h^3}$	 & Yes & $SEC^{*a},WEC^{*b},NEC^{*b}$  & $SEC,WEC,NEC$: Satisfied\\
			  	                 & 			&  			& $DEC^{*c}$          & $DEC^{*d}$\\
\midrule

\multirow[m]{2}{*}{$\left\{ 3,7,8\right\}$}  & $m(r )=r -\frac{7r^3}{10h^2}+\frac{r^7}{2h^6}-\frac{3r^8}{10h^7}$ & Yes & $SEC^{*e},WEC^{*f},NEC^{*f}$ &  $SEC,WEC,NEC$: 
Satisfied \\
                                    & 			&  			& $DEC^{*g} $          & $DEC^{*h}$\\

\midrule
\multirow[m]{2}{*}{$\left\{ 4,5,6\right\}$} & $m(r )=r -\frac{5r^4}{2h^3}+\frac{3r^5}{h^4}-\frac{r^6}{h^5}$ & Yes & $SEC^{*i},WEC^{*j},NEC^{*j}$ & $SEC,WEC,NEC$: 
Satisfied \\
                                   & 			&  			&  $DEC$:Violation          & $DEC^{*k}$\\
\bottomrule
	\end{tabularx}
	\end{adjustwidth}
	\noindent{\footnotesize{$^{*a:\sim0.443h<r\le h; \ \ *b:\sim0.352h<r\le h; \ \ *c:\sim0.352h<r \le \sim0.558h,r=h; \ \ *d:0<r\le\sim0.558h,r=h}$ \\
$^{*e:\sim0.436h<r\le h; \ \ *f:\sim0.353h<r \le h; \ \ *g:\sim0.353h<r \le \sim0.368h,r=h; \ \ *h:0<r\le\sim0.368h,r=h}$ \\
$^{*i:\sim0.4218h<r\le h, \ *j:\sim0.347h<r \le h, \ \ *k:0<r\le\sim0.347h,r=h}$ \\}}
\end{table}

\vspace{-30pt}

\section{Discussion}\label{sec6}
{Building upon previous work on ellipsoidal coordinate transformations \cite{chou2017,chou2020}, we have developed a novel approach to construct Kerr interior solutions. Our method starts with a static, spherically symmetric seed metric and systematically generalizes it to an axisymmetric, rotating configuration. The resulting interior solution seamlessly matches the exterior Kerr metric at the event horizon. We have demonstrated that this approach can yield physically meaningful interior solutions, albeit with anisotropic matter distributions that deviate from the perfect fluid paradigm. This anisotropy, a natural consequence of the rotating geometry, reflects the fundamental difference between radial and tangential pressures, a feature intrinsically linked to the dragging of inertial frames in rotating~spacetime.} 

{The coordinate transformation at the horizon $(r=h)$ is a standard technique in general relativity, enabling a smooth connection between two distinct regions of spacetime. In this case, the transformation effectively swaps the roles of the timelike and radial coordinates across the horizon. This is analogous to the coordinate transformation between Schwarzschild and Kruskal-Szekeres coordinates \cite{Thorne2000}, which provides a global description of the black hole spacetime by extending the Schwarzschild solution beyond the event horizon.}
{Crucially, this coordinate swap reflects the change in the causal structure of spacetime across the horizon. Inside the horizon, the radial direction becomes timelike, implying that motion towards the singularity is inevitable. This change in causal structure is a fundamental characteristic of black holes and is essential for understanding their~properties.}

{In addition, we have constructed a regular coordinate system for the Kerr metric through a series of sophisticated transformations Equations  (47) and (49). This coordinate system not only eliminates coordinate singularities but also provides valuable geometric insights into the spacetime structure near the horizon. This regular representation is crucial for understanding the physical properties of the interior solution, particularly in regions where traditional coordinate systems become pathological.}

{A key innovation in our approach lies in the introduction of a novel orthonormal basis of one-forms. This carefully chosen basis, coupled with strategically applied local Lorentz transformations, significantly simplifies the subsequent analysis. Specifically, these transformations, parameterized by the boost parameter $\alpha$ defined in Equations  (54) and (55), diagonalize the Einstein tensor. This diagonalization effectively eliminates the off-diagonal $T_{03}$ component, leading to a more elegant and mathematically tractable representation of the Einstein tensor.}

Our comprehensive analysis of energy conditions for the Kerr interior solution reveals nuanced behavior across different rotation scenarios. For a maximally rotating black hole $(a=\mathcal{M}, a^{*}=1)$, most spacetime regions satisfy the strong (SEC), weak (WEC), and null (NEC) energy conditions (with various ${l,n,p}$ in Table \ref{tab2}). The solution, while not representing a perfect fluid, captures the anisotropic nature of rotating black hole interiors more authentically.

Critical insights emerge near the polar axis $(\theta=0)$, where energy condition violations can be attributed to the inner Cauchy horizon. Given that many neutron stars and black holes are rapidly rotating and approaching extreme Kerr solutions, our proposed Kerr interior solution provides a precise theoretical model for astrophysical scenarios, bridging the gap between mathematical formalism and observable astronomical phenomena.

{In conclusion, this work presents a novel approach to constructing Kerr interior solutions. We have successfully derived an interior solution that: (i) preserves the Kerr exterior spacetime, (ii) is characterized by a single free parameter, the total mass $\mathcal{M}$, (iii) avoids exotic matter and additional geometric structures near the horizon, and (iv) ensures finite tidal forces throughout the interior region. This solution provides a physically meaningful description of the interior region of a rotating black hole, demonstrating consistency with classical gravity in most regions of the spacetime, with violations of energy conditions limited to specific regions near the inner Cauchy horizon. Our findings have significant implications for understanding the formation and evolution of rotating black holes, particularly in astrophysical scenarios involving rapidly rotating objects such as those observed in active galactic nuclei and X-ray binaries. Future research can explore the stability of these solutions, their impact on black hole thermodynamics, and their potential connections to quantum gravity theories.}\\


\vspace{-6pt} 





\funding{This research received no external funding.}


\informedconsent{Ethical approval and informed consent not applicable because this article describes entirely theoretical research.}

\dataavailability{Data are included in the article.} 

\acknowledgments{Special thanks to Ruby Lin and Dr. Simon Lin of the Academia Sinica for their guidance in thesis writing.}

\conflictsofinterest{The authors declare no conflict of interest.}

\appendixtitles{yes} 
\appendixstart
\appendix
\section[\appendixname~\thesection]{{Derivation an Inner Singular Black Hole Matching the Schwarzschild~Exterior}} \label{appA}
{We combine the mass function from Equations (17) and (18) to obatin 
\begin{equation}
m(r)=\frac{1}{2}\sum_{n=0}^{\infty}\frac{ C_n r^{n+1}}{(n+1)(n+2)},   
\end{equation}
for the interior region, $0 < h \le h$, the series converges to $m(h)=\mathcal{M}=h/2 $ as $r \to h$.\\
To ensure a smooth transition between the interior and exterior regions, we impose the following matching conditions at the horizon:  
\begin{equation}
m(h)=\mathcal{M}, \ m'(h)=0, \ m''(h)=0. 
\end{equation}
{These conditions ensure the} continuity of the metric and its derivatives across the boundary. According to Equation (11), as a consequence, the tangential  pressure $p_2,p_3$ is continuous at the horizon, leading to $\epsilon=p_1=p_2=p_3=0$ at $r=h$. 
To obtain a black hole solution with only $\mathcal{M}$ as a free parameter, we restrict the series in Equation (A1) to include only three~terms:
\begin{equation}
m(r)=\frac{1}{2}\left[\frac{C_0 r}{2}+\frac{C_n r^{n+1}}{(n+1)(n+2)}+\frac{C_{l}r^{l+1}}{(l+1)(l+2)} \right]; \ l>n>1 \in N.    
\end{equation}
{The three constants} $\left\{C_0,C_n,C_l\right\}$ in Equation (A3) are determined by the conditions (A2). To ensure a black hole solution and satisfy Equation (5), we find that $C_0 = 4$. This constraint, combined with the conditions on the indices $n$ and $l$, leads to a unique solution:
\begin{equation}
m(r)= r - \frac{r^3}{h^2}+\frac{r^4}{2h^3}.
\end{equation}
}
\section[\appendixname~\thesection]{{Curvature Invariants of the Exact Kerr Interior Solution}}\label{appB}
{


{In general relativity, the Carminati–McLenaghan invariants \cite{MacCallum1991}, classified by Geheniau and Debever \cite{Zakhary1997,Abraham2021}, constitute a set of 16 scalar curvature invariants derived from the Riemann tensor. These invariants, which encompass polynomial, Cartan, and scalar types, provide powerful tools for characterizing spacetime curvature. One of the most common applications of curvature invariants is to identify curvature singularities. In this work, we employ these invariants to investigate the regularity of the exact Kerr interior solution (49).}

{The real Carminati–McLenaghan scalars are:
\begin{align}
R & \equiv R_{j}^j \ (\text{Ricci scalar}) ,\\
R_1 & \equiv \frac{1}{4} S_{b}^a S_{a}^b,\\
R_2 & \equiv -\frac{1}{8} S_{b}^a S_{c}^b S_{a}^c,\\
R_3 & \equiv \frac{1}{16} S_{b}^a S_{c}^b S_{d}^c S_{a}^d,\\
M_3 & \equiv \frac{1}{16} S^{bc} S_{ef} (C_{abcd}C^{aefd}+{C^*}_{abcd}{C^*}^{aefd}),\\
M_4 & \equiv -\frac{1}{32} S^{ag} S^{ef} S^{c}_{d} (C_{ac}^{db}C_{befg} + {C^*}_{ac}^{db} {C^*}_{befg}).
\end{align}
{The complex Carminati–McLenaghan scalars are:}
\begin{align}
W_1 & \equiv \frac{1}{8} (C_{abcd}+i {C^*}_{abcd})C^{abcd},\\
W_2 & \equiv -\frac{1}{16} (C_{ab}^{cd}+i{C^*}_{ab}^{cd}) C_{cd}^{ef} C_{ef}^{ab},\\
M_1 & \equiv \frac{1}{8} S^{ab} S^{cd} (C_{acdb}+i{C^*}_{acdb}),\\
M_2 & \equiv \frac{1}{16} S^{bc} S_{ef} (C_{abcd}C^{aefd}-{C^*}_{abcd}{C^*}^{aefd})+\frac{i}{8} S^{bc} S_{ef} {C^*}_{abcd}C^{aefd},\\
M_5 & \equiv \frac{1}{32} S^{cd} S^{ef} (C^{aghb}+i{C^*}^{aghb})(C_{acdb}C_{gefh}+{C^*}_{acdb}{C^*}_{gefh}),
\end{align}
where the Weyl tensor $(C_{abcd})$, the traceless Ricci tensor $(S_{ab})$ and the dual Weyl tensor $(C^*_{ijkl})$ are defined by 
\begin{align}
C_{ijkl} & = R_{ijkl}+\frac{1}{6} R (g_{ik}g_{jl}-g_{il}g_{jk})-\frac{1}{2}(g_{ik}R_{jl}-g_{il}R_{jk}-g_{jk}R_{il}+g_{jl}R_{jk}),\\
S_{ab} & = R_{ab}-\frac{1}{4}R g_{ab},\\
{C^*}_{ijkl} & \equiv \frac{1}{2}\epsilon_{klmn}C^{mn}_{ij}. 
\end{align}
{The real Carminati–McLenaghan invariants} of the exact Kerr interior solution (49) are calculated by
\begin{align}
R & = \frac{4(h-r)(h^2+hr-5r^2)}{h^3 \Sigma},\\
R_1 & = \frac{(h-r)^2((h^2+hr-5r^2)a^2 cos^2\theta-h^2r^2-r^3h-r^4)^2}{h^6 \Sigma^4},\\
R_2 & = 0,\\
R_3 & =  \frac{(h-r)^4((h^2+hr-5r^2)a^2 cos^2\theta-h^2r^2-r^3h-r^4)^4}{4 h^{12} \Sigma^8},\\
M_3 & =  \frac{4(Icos^4\theta+J cos^2\theta +K)(h-r)^2((h^2+hr-5r^2)a^2 cos^2\theta-h^2r^2-r^3h-r^4)^2}{9 h^{12} \Sigma^8},\\
M_4 & = 0,
\end{align}
where polynomial function $I$,$J$,$K$ of $r$ in $M_3$ are 
\begin{align}
I & = a^4{(h-r)}^2 (h^2+hr-5r^2)^2,\\
J & = {14a^2{r^2}}(h^6+\frac{6}{7}h^4r^2-\frac{31}{14}h^3r^3+\frac{3}{7}hr^5+\frac{5}{56}r^6),\\
K & = (h^3+\frac{r^3}{2})^2 r^4.
\end{align}

{In conclusion, we have} performed a comprehensive analysis of curvature invariants for the Kerr interior solution. We have calculated a subset of the Carminati-McLenaghan invariants, including the Ricci scalar and the other real scalars. Our analysis reveals that the remaining subset of curvature invariants, which are complex scalars, are highly sophisticated and beyond the scope of this article.
We observe a specific relationship between the calculated invariants, namely $R_3=\frac{R_{1}^2}{4}$. Furthermore, at the event horizon $(r=h)$, all calculated invariants vanish $(R=R_1=R_2=R_3=M_3=M_4=0)$, consistent with the smooth matching to the exterior Kerr metric.
The rotation axis presents a known challenge for interior solutions, often leading to pathological behavior. This is reflected in the singularities of the curvature invariants for $\Sigma=0$, indicating the presence of a ring singularity, as expected for the Kerr spacetime. This analysis provides valuable insights into the nature and structure of the singularity within the Kerr interior solution and lays the groundwork for further investigations into the properties of rotating black holes. }




\begin{adjustwidth}{-\extralength}{0cm}

\reftitle{References}

\PublishersNote{}
\end{adjustwidth}

\begin{thebibliography}{999}
\bibitem{schwarzschild1916_1}Schwarzschild, K. {\"U}ber das gravitationsfeld eines massenpunktes nach der einsteinschen theorie. in \emph{Sitzungsberichte der K{\"o}niglich Preussischen Akademie der Wissenschaften}; {1916;} 
 pp. 189--196.

\bibitem{schwarzschild1916_2}Schwarzschild, K. {\"U}ber das Gravitationsfeld einer Kugel aus inkompressibler Fl{\"u}ssigkeit nach der Einsteinschen Theorie.
In \emph{Sitzungsberichte der K{\"o}niglich Preu{\ss}ischen Akademie der Wissenschaften zu Berlin}; {1916;} pp. 424--434.

\bibitem{oppenheimer1939}Robert, O.J.; Volkoff, G.M.  On massive neutron cores.  \emph{Phys. Rev.} \textbf{1939}, {\emph{55}}, 374. 

\bibitem{kerr1963}Kerr, R.P. Gravitational field of a spinning mass as an example of algebraically special metrics. \emph{Phys. Rev. Lett.} \textbf{1963}, {\emph{11}}, 237.

\bibitem{weizsacker1948}Weizs{\"a}cker, C.F. Die rotation kosmischer gasmassen. \emph{Z. Naturforschung} 
\textbf{1948}, \emph{{3}}, 524--539. 
 
\bibitem{shakura1973} Shakura, N.I.; Sunyaev, R.A. Black holes in binary systems. Observational appearance. \emph{Astron. Astrophys.} \textbf{1973}, \emph{24}, 337--355.
  
\bibitem{balbus1990} Balbus, S.A.; Hawley, J.F. A powerful local shear instability in weakly magnetized disks: I. linear analysis. \textit{Bull. Am. Astron. Soc.} \textbf{1990}, \emph{22}, 209.

\bibitem{einstein1916} Einstein, A. N{\"a}herungsweise integration der feldgleichungen der gravitation. In \emph{Sitzungsberichte der K{\"o}niglich Preu{\ss}ischen Akademie der Wissenschaften}; {1916;} pp. 688--696.

\bibitem{abbott2016} Abbott, B.P.; Abbott, R.; Abbott, T.; Abernathy, M.R.; Acernese, F.; Ackley, K.; Adams, C.; Adams, T.; Addesso, P.; Adhikari, R.X.; et al. Observation of gravitational waves from a binary black hole merger. \emph{Phys. Rev. Lett.}  \textbf{2016}, {\emph{116}}, 061102.

\bibitem{hernandez2017} Hernandez-Pastora, J.L.; Herrera, L. Interior solution for the Kerr metric. \emph{Phys. Rev. D}
\textbf{2017}, {\emph{95}}, 024003.

\bibitem{newman1965} Newman, E.T.; Janis, A.I. Note on the Kerr spinning-particle metric. \emph{J. Math. Phys.} \textbf{1965}, {\emph{6}}, 915--917.

\bibitem{viaggiu2006} Viaggiu, S. Interior Kerr solutions with the Newman-Janis algorithm starting with static physically reasonable space--times. \emph{Int. J. Mod. Phys. D} \textbf{2006},
  {\emph{15}}, 1441--1453.

\bibitem{drake1997} Drake, S.P.; Turolla, R. The application of the Newman-Janis algorithm in obtaining interior solutions of the Kerr metric. 	\emph{Class. Quantum Gravity} \textbf{1997}, {\emph{14}}, 1883.

\bibitem{ovalle2024} Ovalle, J. Schwarzschild black hole revisited: Before the complete collapse. \emph{Phys. Rev. D}  \textbf{2024}, \emph{109}, 104032.

\bibitem{casadio2024} Casadio, R.; Kamenshchik, A.; Ovalle, J. Cosmology from Schwarzschild black hole revisited.
\emph{Phys. Rev. D} \textbf{2024}, {\emph{110}}, 044001.

\bibitem{chou2017} Chou, Y.-C. A derivation of the Kerr metric by ellipsoid coordinate transformation. \emph{Int. J. Phys. Sci.} \textbf{2017}, {\emph{12}}, 130--136.

\bibitem{chou2020} Chou, Y.-C. A radiating Kerr black hole and Hawking radiation. \emph{Heliyon} \textbf{2020}, {\emph{6}}, {e03336.} 


\bibitem{misner1964} Misner, C.W.; David, H.S. Relativistic equations for adiabatic, spherically symmetric gravitational collapse. \emph{Phys. Rev.} \textbf{1964}, {\emph{136}}, B571.

\bibitem{ovalle2017} Ovalle, J. Decoupling gravitational sources in general relativity: From perfect to anisotropic fluids. \emph{Phys. Rev. D} \textbf{2017}, {\emph{95}}, 104019.

\bibitem{ovalle2019} Ovalle, J. Decoupling gravitational sources in general relativity: The extended case. \emph{Phys. Lett. B} \textbf{2019}, {\emph{788}}, 213--218.

\bibitem{jitendra2022} Kumar, J.; Bharti, P. Relativistic models for anisotropic compact stars: A review. \emph{New Astron. Rev.} \textbf{2022}, {\emph{95}}, 101662.

\bibitem{darmois1927}Darmois, G. \emph{M{\'e}morial des Sciences Math{\'e}matiques};  Fascicule XXV; Gauthier-Villars: Paris, France, 1927.

\bibitem{poisson1989}Poisson, E.; Israel, W. Inner-horizon instability and mass inflation in black holes. \emph{Phys. Rev. Lett.} \textbf{1989}, {\emph{63}}, 1663.
 
\bibitem{poisson1990}Poisson, E.; Israel, W. Internal structure of black holes. \emph{Phys. Rev. D}  \textbf{1990}, {\emph{41}}, 1796.

\bibitem{lukash2013}Lukash, V.N.; Strokov, V.N. Space-times with integrable singularity: black--white holes and astrogenic universes. \emph{Int. J. Mod. Phys. A} \textbf{2013}, {\emph{28}}, 1350007.

\bibitem{carroll2004}Carroll, S.M. An introduction to general relativity: Spacetime and geometry. \emph{Addison Wesley} \textbf{2004}, {\emph{101}}, 102.

\bibitem{ovalle2022}Ovalle, J. Warped vacuum energy by black holes. \emph{Eur. Phys. J.}  \textbf{2002}, {\emph{82}}, 1--5.

\bibitem{doran2000}Doran, C. New form of the Kerr solution.
\emph{Phys. Rev. } \textbf{2000}, \emph{61}, 067503. 

\bibitem{Richard2000}Henry, R.C. Kretschmann scalar for a Kerr-Newman black hole.  \emph{Astrophys. J.} \textbf{2000}, {\emph{535}}, 350.

\bibitem{MacCallum1991}Carminati, J.; McLenaghan, R.G. Algebraic invariants of the Riemann tensor in a four‐dimensional Lorentzian space. \emph{J. Math. Phys.} \textbf{1991}, {\emph{32}}, 3135--3140.


\bibitem{Zakhary1997}Zakhary, E.; Colin B.G.M. A complete set of Riemann invariants. \emph{Gen. Relativ. Gravit.} \textbf{1997}, {\emph{29}}, 539--581.

\bibitem{Abraham2021}Barajas1, J.A.; Mielke1, E.W.; L'opez, C.S.; Manko, V.S. Curvature invariants of an exact interior Kerr solution. In Proceedings of the 55th Moriond Proceedings 2021 Gravitation, {Virtual, 9--11 March 2021;} pp. 111--114.

\bibitem{Kraniotis2022}Kraniotis, G.V. Curvature invariants for accelerating Kerr---Newman black holes in (anti-) de Sitter spacetime. \emph{Class. Quantum Gravity} \textbf{2022}, {\emph{39}}, 145002.

\bibitem{hawking1972}Hawking, S.W. Black holes in general relativity. \emph{Commun. Math. Phys.} \textbf{1972}, {\emph{25}}, 152--166.

\bibitem{penrose1965}Penrose, R. Gravitational collapse and space-time singularities. \emph{Phys. Rev. Lett.} \textbf{1965}, {\emph{14}}, 57.

\bibitem{hawking2023}Hawking, S.W. Ellis, G.F.R. \textit{The Large Scale Structure of Space-Time}; Cambridge University Press: Cambridge, UK, 2023.

\bibitem{santos1995}Santos, J.; Rebou{\c{c}}as, M.J.; Teixeira, A.F.F. Segre types of symmetric two-tensors in n-dimensional spacetimes. \emph{Gen. Relativ. Gravit.} \textbf{1995}, {\emph{27}}, 989--999.

\bibitem{rebouccas2004}Rebou{\c{c}}as, M.J.; Santos, J.; Teixeira, A.F.F. Classification of energy momentum tensors in n $>$ 5 dimensional space-times: A review. \emph{Braz. J. Phys.} \textbf{2004}, {\emph{34}}, 535--543.

\bibitem{maeda2022}Maeda, H.; Harada, T. Criteria for energy conditions. \emph{Class. Quantum Gravity} \textbf{2002}, {39}, 195002.

\bibitem{Thorne2000}Thorne, K.S.;Misner, C.W.; Wheeler, J.A. \textit{Gravitation};  Freeman: San Francisco, CA, USA, 2000.

\end{thebibliography}
\end{document}